\input phyzzx.tex
\tolerance=1000
\voffset=-0.0cm
\hoffset=0.7cm
\sequentialequations
\def\rl{\rightline}

\def\t1{{\tilde 1}}

\def\t{\theta}

\REF{\COB}{C. L. Bennett et. al. astro-ph/9601067; C. B. Netterfield et. al. astro-ph/0104460; A. T. Lee et. al. astro-ph/0104459.}
\REF{\GUT}{A. H. Guth, Phys. Rev {\bf D23} (1981) 347.}
\REF{\LIN}{A. D. Linde, Phys. Lett. {\bf B108} (1982) 389.}
\REF{\ALB}{A. Albrecht and P. J. Steinhardt, Phys. Rev. Lett. {\bf 48} (1982) 1220.}
\REF{\RAP}{R. Bousso, hep-th/0203101 and references therein.}
\REF{\STR}{A. Strominger, JHEP {\bf 0110} (22001) 034; hep-th/0106113.}
\REF{\SSV}{M. Spradlin, A. Strominger and A. Volovich, hep-th/0110007.}
\REF{\DAN}{U. H. Danielsson, hep-th/0110265.}
\REF{\ADS}{E. Halyo, hep-th/0112093.}
\REF{\KLE}{D. Klemm, hep-th/0106247.}
\REF{\NO}{S. Nojiri and S. D. Odintsov, hep-th/0106191; hep-th/0107134; hep-th/0112152.}
\REF{\DES}{E. Halyo, hep-th/0107169.}
\REF{\INF}{A. Strominger, hep-th/0110087.}
\REF{\SIT}{T. Shiromizu, D. Ita and T. Torii, hep-th/0109057.}
\REF{\CAI}{R. G. Cai, hep-th/0111093; hep-th/0112253.}
\REF{\MED}{A. J. M. Medved, hep-th/0111182; hep-th/0111238; hep-th/0112226.}
\REF{\OGU}{S. Ogushi, hep-th/0111008.}
\REF{\BBM}{V. Balasubramanian, J. de Boer and D. Minic, hep-th/0110108.}
\REF{\PET}{A. C. Petkou and G. Siopsis, hep-th/0111085.}
\REF{\MAN}{A. M. Ghezelbash and R. B. Mann, hep-th/0111217; hep-th/0201004.}
\REF{\CVE}{M. Cvetic, S. Nojiri and S. D. Odintsov, hep-th/0112045.}
\REF{\BMS}{R. Bousso, A. Maloney and A. Strominger, hep-th/0112218.}
\REF{\SPR}{M. Spradlin and A. Volovich, hep-th/0112223.}
\REF{\STR}{E. Halyo, hep-th/0201174.}
\REF{\GIL}{D. Kabat and G. Lifschytz, hep-th/0203083.}
\REF{\LAR}{F. Larsen, J. P. van der Saar and R. G. Leigh; hep-th/0202127.}
\REF{\RIO}{D. H. Lyth and A. Riotto, Phys. Rep. {\bf 314} (1999) 1, hep-ph/9807278.}
\REF{\LEN}{L. Dyson, J. Lindesay and L. Susskind, hep-th/0202163.}
\REF{\CHA}{A. D. Linde, Phys. Lett. {\bf B129} (1983) 177; Phys. Lett. {\bf B175} (1986) 395.}
\REF{\QUI}{B. Ratra and P. J. E. Peebles, Phys. Rev. {\bf D37} (1988) 3406.}
\REF{\STE}{L. Wang, R. R. Caldwell, J. P. Ostreiker and P. J. Steinhardt, Astrophys. {\bf J530} (2000) 17, astro-ph/9901388.}
\REF{\HYB}{A. D. Linde, Phys. Lett. {\bf B259} (1991) 38; Phys. Rev. {\bf D49} (1994) 748.}
\REF{\DTE}{E. Halyo, Phys. Lett. {\bf B387} (1996) 43, hep-ph/9606423.}
\REF{\BIN}{P. Binetruy and G. Dvali, Phys. Lett. {\bf B450} (1996) 241, hep-ph/9606342.}
\REF{\TYP}{E. Halyo, Phys. Lett. {\bf B454} (1999) 223, hep-ph/9901302.}
\REF{\REN}{R. Kallosh, hep-th/0109168.}
\REF{\MAR}{R. Kallosh, A. Linde, S. Prokushkin and M. Shmakova, hep-th/0110089.}
\REF{\BRA}{G. Dvali and S.-H. H. Tye, Phys. Lett. {\bf B450} (1999) 72, hep-th/9912483.}
\REF{\DSS}{G. Dvali, Q. Shafi and S. Solganik, hep-th/0105203.}
\REF{\BUR}{C. P. Burgess at. al. JHEP {\bf 07} (2001) 047, hep-th/0105204.}
\REF{\ALE}{S. H. Alexander, Phys. Rev {\bf D65} (2002) 023507, hep-th/0105032.}
\REF{\EDI}{E. Halyo, hep-ph/0105216; hep-ph/0105341.}
\REF{\SHI}{G. Shiu and S.-H. H. Tye, Phys. Lett. {\bf B516} (2001) 421, hep-th/0106274.}
\REF{\CAR}{C. Herdeiro, S. Hirano and R. Kallosh, JHEP {\bf 0112} (2001) 027, hep-th/0110271.}
\REF{\SHA}{B. S. Kyae and Q. Shafi, Phys. Lett. {\bf B526} (2002) 379, hep-ph/0111101.}
\REF{\BEL}{J. Garcia-Bellido, R. Rabadan and F. Zamora, JHEP {\bf 01} (2002) 036, hep-th/0112147.}
\REF{\KAL}{K. Dasgupta, C. Herdeiro, S. Hirano and R. Kallosh, hep-th/0203019.}

\singlespace
\rl{SU-ITP-01-03}
\rl{hep-ph/0203235}
\rl{\today}
\pagenumber=0
\normalspace
\medskip
\bigskip
\titlestyle{\bf{Holographic Inflation}}
\smallskip
\author{ Edi Halyo{\footnote*{e--mail address: vhalyo@stanford.edu}}}
\smallskip
\centerline {Department of Physics}
\centerline{Stanford University}
\centerline {Stanford, CA 94305}
\centerline{and}
 \centerline{California Institute for Physics and Astrophysics}
\centerline{366 Cambridge St.}
\centerline{Palo Alto, CA 94306}
\smallskip
\vskip 2 cm
\titlestyle{\bf ABSTRACT}

Using the de Sitter/CFT correspondence we describe a scenario of holographic inflation which is driven by a three dimensional boundary field theory.
We find that inflationary constraints severely restrict the $\beta$--function, the anomalous dimensions and the value of the $C$--function of the boundary theory.
The scenario has model independent predictions such as $\epsilon<< \eta$, $n_T<0.04$, $P_{tensor}/P_{scalar}<0.08$ and $H<10^{14}~GeV$. We consider some simple
boundary theories and find that they do not lead to inflation. Thus, building an acceptable holographic inflation model remains a challenge.
We also describe holographic quintessence and find that it closely resembles a cosmological constant.

\singlespace
\vskip 0.5cm
\endpage
\normalspace

\centerline{\bf 1. Introduction}
\medskip

Inflation is the only paradigm for the early universe that agrees with a large number of cosmological observations (such as the
COBE and BOOMERANG data etc.[\COB]) and solves the problems associated with the Big Bang cosmology[\GUT,\LIN,\ALB]. Thus it is quite important to build models of
inflation especially those with predictions that can be tested in the future. On the other hand, inflation is a gravitational
phenomenon and therefore should be related to the holographic principle which is believed to be one of the fundamental principles of the true theory of gravity[\RAP].
Therefore, we expect that inflation could in principle be described holographically, i.e. in terms
of a three dimensional field theory that lives on the boundary of the (inflating) universe.

Recently, a holographic duality between de Sitter space--time and a CFT living on the boundary of the space was conjectured[\STR]. There is a large amount of circumstantial
evidence for this conjecture[\STR-\GIL]; however it is also plagued by some serious problems such as the nonunitarity of the boundary field theory and the lack of a definition of time
evolution on the boundary. In this paper we simply assume that the conjecture of the dS/CFT correspondence is correct.
According to this correspondence a massive
scalar field in the bulk corresponds to an operator in the boundary theory with an anomalous dimension (proportional to the bulk mass for small masses). In addition,
time evolution in the bulk corresponds to an RG flow from the IR to the UV on the boundary[\INF]. The radius of the de Sitter space which is related to the vacuum
energy is inversely proportional to the central charge of the boundary CFT. In [\DAN,\ADS] it was shown that the dS/CFT correspondence can be generalized to asymptotically
de Sitter space--times. In these
cases the boundary theories are field theories with nonzero $\beta$--functions which flow to a fixed point.

A well--known example of a nearly de Sitter space is the universe during inflation. Thus the dS/CFT correspondence can be used to analyze and possibly build
new, holographic models of
inflation. In this context, the inflating universe in four space--time dimensions is holographically dual to a boundary field theory in three Euclidean dimensions.
The inflaton field in the bulk, $\phi$ corresponds to an operator $O$ in the boundary theory.
The nonzero inflaton potential is related to the (finite) value of the $C$--function of the boundary theory. The slow--roll of the inflaton down its
potential is described by a small $\beta$--function. The very early times during which the universe undergoes inflation correspond to the far IR on the
boundary.

In fact one can make this correspondence more precise.
Recently it was shown that the slow--roll parameters of inflation $\epsilon$ and $\eta$ are completely fixed by the $\beta$--function and $\gamma_O$,the anomalous dimension of
$O$[\LAR].
Then slow--roll inflation simply requires $\beta,\gamma<<1$ on the boundary field theory.
However a successful model of inflation should also give at least 60 e--foldings and produce the correct magnitude of density perturbations. In this paper we translate
these constraints into the language of the boundary theory and show that they strongly constrain the possible $\beta$--functions and values of the $C$--function.
The only constraint on the anomalous dimension of $O$ comes from the slow--roll condition. We show that in principle these constraints can be satisfied by three dimensional
Euclidean boundary field theories which would constitute models of holographic inflation. However in practice this turns out to be more difficult; we give a couple of examples of boundary
field theories and show why they fail to produce acceptable models of inflation in the bulk. Construction of a working model of holographic inflation remains a challenge.

We find that the holographic inflation scenario has some model independent predictions. First, if we demand that the boundary theory remain perturbative near the IR fixed
point (in order to be calculable) we find $H < 10^{14}~GeV$ during inflation.
It is amusing that this is exactly the upper bound obtained from COBE data. Second, combining this with the observed magnitude of density perturbations gives the upper bounds
$\epsilon<0.02$, $P_{tensor}/P_{scalar}<0.08$ and $|n_T|<0.04$. Thus holographic inflation predicts a nearly scale invariant tensor component of density perturbations
which is more than an order of magnitude smaller than that of the scalar component. The tilt (scalar index), $n_S$, depends on the
anomalous dimension of $O$ and therefore is much less restricted. However, another robust prediction of the scenario is $\epsilon<< \eta$ which means that the deviation from
scale invariance is much larger for scalar perturbations than for tensor perturbations.
Third, the bounds on $\epsilon$, $n_T$ and $P_{tensor}/P_{scalar}$ are proportional to $H^2/M_P^2$; thus
the lower the scale of inflation the stronger the bounds.

Models of holographic inflation correspond to bulk scalar potentials which cannot be obtained by other methods. The form of the potential is exponential but not
derivable in e.g. supergravity; moreover the bulk potential contains small numerical factors which look like fine tuning. On the other hand, from the boundary point of view these
small numbers are very natural and arise as one--loop coefficients of $\beta$ and $\gamma$ functions. It is precisely these small numbers that allow for
acceptable inflation that satisfies all constraints. Thus holography through the dS/CFT correspondence seems to generate and stabilize these unnatural
potentials which lead to inflation. We also show that the same ideas apply to holographic quintessence. However, in this case the observational bounds can only
reproduce the slow--roll condition. The only prediction of holographic quintessence is (assuming a perturbative boundary theory) an $\omega$ very close to $-1$, i.e.
a behavior very close to a cosmological constant.

The paper is organized as follows. In the next section we describe the basics of holographic inflation including all the inflationary constraints on the
boundary field theory. In section 3 we describe a couple of possible scenarios with examples and discuss the corresponding bulk potentials. In section 4 we describe
holographic quintessence. Section 5 contains our conclusions and a discussion of our results.

\medskip
\centerline{\bf 2. Holographic Inflation}
\medskip

In this section we describe the holographic inflation scenario which utilizes the holographic field theory dual of the inflating universe to drive inflation
instead of the usual inflaton potential.

Consider the universe with the Robertson--Walker metric
$$ds^2=-dt^2+a(t)^2(dr^2+r^2d \Omega_2^2) \eqno(1)$$
Inflation corresponds to an almost time--independent Hubble constant $H=\dot a(t)/a(t)$ which is equivalent to an exponential scale factor $a(t) \sim e^{Ht}$.
In its simplest version, inflation can be realized by the nonzero vacuum energy of a single scalar inflaton field with potential $V(\phi)$ and equation of motion
$$\ddot \phi +3H \dot \phi +{dV \over d\phi}=0 \eqno(2)$$
where $H^2= V/3 M_P^2$.
In order to have (slow--roll) inflation, $V(\phi)$ has to be very flat. More concretely, $V(\phi)$
has to satisfy the slow--roll conditions
$$\epsilon={1 \over 2}M_P^2 ({V^{\prime} \over V})^2<<1 \qquad \eta=M_P^2{V^{\prime \prime} \over V}<<1 \eqno(3)$$
The above conditions severely constrain the form of the scalar potential $V(\phi)$. Another constraint on $V(\phi)$ arises from the requirement for
(at least) 60 e-foldings during inflation
$$N(\phi)=\int^{\phi_i}_{\phi_c} M_P^{-2} {V(\phi) \over V^{\prime}(\phi)} d\phi \sim 60 \eqno(4)$$
Here $\phi_i$ is the initial value of the inflaton whereas $\phi_c$ is its critical value determined by the end of slow--roll (or inflation) which occurs
when $\epsilon(\phi_c), \eta(\phi_c) \sim 1$.
In addition, the inflaton potential should also satisfy the constraint arising from the COBE observations of the density perturbations
$${\delta \rho \over \rho} \sim {1 \over {5 \pi \sqrt 3}} {V(\phi)^{3/2} \over {M_P^3 V^{\prime}(\phi)}} \sim 2 \times 10^{-5} \eqno(5) $$
One can also measure the power spectrum of the scalar component of the density perturbations arising from the quantum fluctuations ($\delta \phi \sim H/2\pi$)
of the inflaton
$$P_{scalar}=\left ({H \over {\dot \phi}} \right ) \left ({H \over {2 \pi}} \right )^2 \sim k^{n_S-1} \eqno(6)$$
The tilt of the spectrum is given by $n_S=1-4 \epsilon+2 \eta$. The scalar component of the density perturbations is not scale invariant due to the (very weak)
time dependence of the inflaton potential. The deviation from scale invariance is parametrized by $n_S-1$. Similarly one
can define the power spectrum of the tensor component of density perturbations arising from fluctuations in the graviton
$$P_{tensor}={1 \over {4 \pi M_P^2}} \left ({H \over {2 \pi}} \right )^2 \sim k^{n_T} \eqno(7)$$
where $n_T=-2 \epsilon$. Finally one can measure the ratio of powers of the scalar and tensor components
$$r={P_{tensor} \over P_{scalar}}=4 \epsilon  \eqno(8)$$
whose value is a signature of a particular model of inflation.

The aim of inflationary model building is to find a scalar (inflaton) $\phi$ with a suitable potential $V(\phi)$ which satisfies the four constraints in eqs. (3)-(5).
Then, such a model would make predictions about $P_{scalar}, n_S, P_{tensor}$ and $n_T$ which can be checked by observations. In the last twenty years every possible
microscopic potential derived from
models of high energy physics has been used for inflation[\RIO]. A property common to all of these trials is the fact that inflation is driven by scalar fields that live
in the four dimensional inflating universe.
On the other hand, nowadays it is widely believed that holography is a fundamental priciple of the true theory of quantum gravity[\RAP].
According to the holographic principle, the inflationary bulk (gravitational) physics should be described by
a nongravitational theory on the boundary of the space--time.

The de Sitter/CFT correspondence[\STR] which was recently proposed is precisely what is needed for a holographic
description of inflation.
According to the dS/CFT correspondence physics in the de Sitter bulk can be described by an Euclidean CFT on the (future or past) boundary of de Sitter space.
In addition to the asymptotic symmetries of the space there is a large amount of circumstantial evidence in favor of this duality. (For a critisism of the correspondence
see [\LEN].) Thus, in this context the four
dimensional de Sitter space is described by a three dimensional Euclidean field theory at a fixed point.
However, inflation does not exactly correspond to de Sitter space since it eventually ends whereas de Sitter space is eternal. In other words, during inflation
the Hubble constant is not time--independent but changes slowly. In the boundary theory, this corresponds to a perturbation of the CFT which takes the theory away from
the fixed point. For any inflaton field $\phi$ with asymptotic value $\phi_0(x)$ at the boundary ($ t \to + \infty$ or $t \to - \infty$) there is a dual operator $O$ in the boundary
theory such that the boundary theory is described by
$${\cal L}={\cal L}_{CFT}+gO \eqno(9)$$
where the coupling of the perturbation is $g=\phi_0/ M_P$.
It is easy to see from the metric in eq. (1) that time evolution in the bulk is dual to scale transformations on the boundary with the boundary scale  $\mu \sim a(t)$.
We see that early (late) times in the bulk correspond to the IR (UV) on the boundary.
Thus the slow--roll of the inflaton field is described on the boundary by an RG flow from the IR to the UV described by a small (i.e. $<<1$) $\beta$-function[\LAR]
$$\beta={\partial g \over {\partial log \mu}}={1 \over { M_P^2}}{\partial \phi \over {\partial log a}}=-{{2 M_P^2} \over H}{dH \over d \phi} \eqno(10)$$
When $H$ depends weakly on $\phi$ we have a slowly rolling scalar which corresponds on the boundary to a theory with $\beta<<1$.
As a result of this RG run the operator $O$ obtains an anomalous dimension given by[\STR,\LAR]
$$\gamma_O=-{3 \over 2}+\sqrt{(3/2)^2-(m/H)^2} \eqno(11)$$
where $m^2=d^2V(\phi)/d\phi^2$ is the mass of the inflaton. We see that a small inflaton mass corresponds to $\gamma_O<<1$ for the boundary theory.
Finally the $C$--function of the boundary theory is given by[\INF,\ADS,\LAR]
$$C={3M_P^2 \over H^2}={9M_P^4 \over V} \eqno(12)$$
Note that as time goes on the inflaton rolls down its potential, i.e. $V(\phi)$ decreases with time. According to eq. (12), on the boundary this corresponds to an
increasing $C$--function as we flow from the IR to the UV which is exactly the expected behavior.
We see that the $C$--function of the boundary theory parametrizes (the inverse of) the energy density during inflation in Planck units. COBE data already provides the bound
$H<10^{14}~GeV$ which by using eq. (12) gives $C>10^8$. This is
quite surprising especially since inflation corresponds to the far IR limit of the boundary theory which is not expected to have too many degrees of freedom.
On the other hand, $C$ parametrizes the inverse of the inflaton energy density in Planck units. Therefore, if we want to have inflation at an energy scale
below $M_P$ a large value for $C$ is unavoidable.

Now we would like to find the conditions on the boundary theory which will lead to acceptable inflation in the bulk. For this purpose we need to express the conditions
for inflation, in eqs. (3)-(5) in terms of the parameters of the boundary theory. In ref. [\LAR] it was shown that
$$\epsilon=-{\partial log H \over {\partial log a}}=2 M_P^2 \left( {1 \over H} {dH \over {d \phi}} \right)^2 \eqno(13)$$
and
$$\eta=-{\partial log (\partial H/\partial \phi) \over {\partial log a}}=2M_P^2 {1 \over H}{d^2H \over d \phi^2} \eqno(14)$$
Using eqs. (10-11) and (13-14) we find that
$$\epsilon={1 \over 2} \beta^2 \qquad \eta=-\gamma_O \eqno(15)$$
Thus the slow--roll conditions in eq. (3) correspond to $\beta<<1$ and $\gamma_O<<1$ in the boundary theory.

However this is not enough; we also need to make sure that we get at least 60 e--foldings and the correct amount of density perturbations from inflation.
Using eqs. (4) and (10) we find that
$$N(\phi)=\int^{g_i}_{g_c} {1 \over \beta} dg  \sim 60 \eqno(16)$$
where $g_i$ ($g_c$) is the initial (critical) value of the coupling during inflation corresponding to $\phi_i$ ($\phi_c$).
This condition severely constrains the coefficient of the $\beta$--function as
we will see in the next section.

The constraint coming from the density perturbations becomes using eqs. (5) and (14)
$${\delta \rho \over \rho} \sim {1 \over 25}{1 \over {\beta \sqrt c}} \sim 2 \times 10^{-5} \eqno(17)$$
Above we saw that COBE data requires $C>10^8$. Using this constraint in eq. (17) we find that $\beta<1/5$ which satisfies the requirement of slow--roll inflation.
This in turn results in the upper bound
$\epsilon<0.02$ as a prediction of holographic inflation.
Note that the two constraints in eqs. (16) and (17) involve only $\beta$ and $C$ but not $\gamma_O$ which is constrained only by the slow--roll condition.

One can also show that in terms of the boundary theory the scalar and tensor components of density perturbations are described by[\LAR]
$$n_S=1-2 \beta^2-2\gamma_O \qquad n_T=-\beta^2 \eqno(18)$$
Again using $\beta<1/5$ we find $|n_T|<0.04$ giving an almost scale invariant tensor component of density perturbations.
In addition, eq. (8) gives $r<0.08$ which means that the power of the scalar perturbations are at least an order of magnitude larger than those of tensor perturbations.
Using eqs. (15) and (17) we find that $\epsilon \sim 10^6 \times (H^2/M_P^2)$. Therefore, $n_T$ and $r$ are proportional to the value of the inflaton potential resulting in
stronger bounds for lower inflation scales.
The tilt $n_S$ is not constrained in a model independent manner since it depends on $\gamma_O$ which does not appear in eqs. (16) and (17).

Slow--roll inflation ends when the slow--roll conditions cease to hold, i.e. when $\beta,\gamma \sim 1$. Thus we expect that during the RG flow from the IR to the UV, at
some energy scale the $\beta$--function gets large which signals the end of holographic inflation. After this point, the inflaton rolls down its potential fast and
reaches the true, stable minimum of the potential. On the boundary this will corrrespond to a large increase in the value of the $C$--function as the energy scale changes.
Usually, the true, stable minimum of the inflaton potential has vanishing vacuum energy which on the boundary translates to a theory with diverging central charge.
Whether the inflaton oscillates a few or many times about the minimum depends on the ratio $m/H$ near the true minimum. On the boundary this ratio is given by $\gamma$ near the
UV fixed point. Another possibility is for quintessence to follow inflation at very late times. In this case after inflation the vacuum energy is very small but
nonvanishing which describes quintessence. On the boundary one will see an RG flow from an IR fixed point towards a UV fixed point with a much larger but finite central charge.

\medskip
\centerline{\bf 3. Inflationary Boundary Field Theories}
\medskip

In this section, we consider the constraints arising from inflation on the boundary field theory in terms of $\beta$, $\gamma_O$ and $C$ and find that
they are in principle easy to satisfy. However, in practice, finding explicit three dimensional field theoretical models that lead to acceptable inflation seems to be hard.
The consruction of a working model of haolographic inflation remains a challenge.

For example, assume that the boundary theory in the far IR limit has $\beta(g) \sim bg^2$. Then from eq. (16) we find that 60 e--foldings require $bg_i \sim 1/60$.
Assuming that this condition is satisfied, the correct magnitude of density perturbations require $\sqrt c g_i \sim 10^5$. In principle a theory with
$b \sim 10^{-1}$ and $C \sim 10^{12}$ (and $g_i \sim 1/6$) easily satisfies all these constraints. Note that $b$ has to be somewhat larger than $\sim 1/50$ in order for the field
theory to be in the perturbative region. In addition we must assume that $\gamma_O<<1$ since we cannot say much about $\gamma_O$ without explicit knowledge
of the boundary theory. Then a boundary field theory with the above parameters predicts
$n_S \sim 1-2 \gamma_O$ and $n_T \sim - 10^{-5}$. For the type of $\beta$--function above and assuming that $\gamma \sim g$ we get $\epsilon<<\eta$ for weak coupling.
We find that generically the tensor component of density perturbations is extremely close to being scale invariant. In addition, we find $r \sim 6 \times 10^{-6}$ which
means that the tensor component of the density perturbations is negligible compared to the of the scalar component.
Note that the boundary theory has a very large value for the $C$--function $C \sim 10^{12}$ which only reflects the energy scale of inflation, i.e. $V^{1/4} \sim 10^{16}~GeV$.

As an example of a theory of the above type consider the $\phi^6$ theory in $D=3$ with
$${\cal L}={1 \over 2} (\partial_{\mu} \phi)^2+{g \over {6!}} \phi^6 \eqno(19)$$
Here the kinetic term gives ${\cal L}_{CFT}$ and the interaction term is $gO$ in eq. (9). The interaction is marginally irrelevant at the quantum level. In this case
one obtains from the one--loop perturbation theory
$$\beta={5 \over 3} {g^2 \over {16 \pi^2}} \qquad \gamma_{\phi^6}={5 \over 3}{g \over {16 \pi^2}} \eqno(20)$$
As noted in ref. [] these easily satisfy $\beta<<1$ and $\gamma_O<<1$ in the perturbative regime with $g<1$. However, in order to get 60 e--foldings
one needs $g_i \sim 1.6$ which means that there cannot be enough e--foldings
in the perturbative theory. On the other hand, the $\beta$ and $\gamma_O$ are calculated in perturbation theory and therefore this model is not useful for inflation. This
example illustrates the importance of the constraints coming from the number of e--foldings and density perturbations.

Another possibility is a field theory with $\beta(g) \sim ag-bg^2$. This kind of theory has two fixed points; an IR fixed point at $g=0$ and a UV fixed point
at $g=a/b$. In the far IR
we are near $g=0$ and therefore $\beta \sim ag$. Then, from eq. (16) we find that in order to get enough e--foldings we need $g_i \sim e^{-60a}$. COBE data can be reproduced
if in addition $C a^2 g_i^2 \sim 10^{10}$ is also satisfied. All of the above constraints can be satisifed with $a \sim 10^{-1}$ and $C \sim 10^{15}$ (and $g_i \sim e^{-6}$).
Again one needs to assume that $\gamma_O<<1$. In this case, even though $\beta,\gamma \sim g$ we still find $\epsilon << \eta$.
We find that $n_T \sim 10^{-8}$ and $r \sim 4 \times 10^{-8}$ which means that the tensor component of density perturbations is again negligible compared to the
scalar component and extremely close to being scale invariant. The very large value of $C \sim 10^{15}$ again reflects the energy scale of inflation given by
$V^{1/4} \sim 10^{15}~GeV$.

As an example of a field theory of the type above, consider the nonlinear $\sigma$ model in $D=3$ with
$${\cal L}={1 \over {2g^2}}(\partial_{\mu} \phi_i)^2 \eqno(21)$$
where $\phi_i^2=1$. Since $g$ is dimensionful we define a dimensionless coupling $T=g^2 M$. This theory has at the one--loop level
$$\beta=T-(N-2){T^2 \over {2 \pi}} \qquad  \gamma=(N-1) {T \over {4 \pi}}       \eqno(22)$$
We see that there is an IR fixed point at $T=0$ and a UV fixed point at $T=2 \pi/(N-2)$. The slow--roll conditions near $T=0$ require that $T_i<<1$ and $N T_i \sim 1$.
From the previous discussion we find that in order to get enough e--foldings we need $g_i \sim 10^{-30} M^{-1/2}$. The conditon in eq. (17) becomes $C \sim 10^{60}$ which means that
the energy scale at which inflation occurs is very low; i.e TeV scale which leads to many problems

We can reconstruct the bulk potential of the inflaton during inflation from the boundary data. Consider a boundary field theory with $\beta \sim ag^n$ in the far IR
(for $g<<1$) with $a<<1$. Now using eq. (10) and $g=\phi/M_P$ we find
$$V(\phi)=H_0^2 M_P^2  exp \left( -{{2a} \over (n+1)}{\phi^{n+1} \over M_P^{n+1}} \right) \eqno(23)$$
where $H_0$ is an integration constant to be determined. Since holographic inflation takes place in the perturbative regime with $g_i<<1$ and $c=3M_P^4/V>10^8$ we find
$H_0 < 10^{-4} M_P$ which agrees with the upper bound from COBE data. We want to stress that this is the shape of the potential that corresponds only to the inflationary era.
Slow--roll inflation ends when $\beta \sim 1$
and then higher order terms in the expansion of $\beta$ become important and the function $\beta(g)$ takes a complicated form.
Eq. (23) describes an exponential potential which has been known to be problematic for inflation for a long time. This kind of an inflaton potential with $n=1$ can be obtained
from supergravity. (For $n>1$ the potential cannot be obtained in supergravity.)
The scalar supergravity potential looks similar to eq. (23) if the scalar has a canonical Kahler potential. In this case it is well--known that the inflaton
mass is about the Hubble constant and therefore slow--roll inflation cannot take place. On the other hand, in the above potential the exponent in eq. (23) contains a small numerical
factor, $a$. This is completely unnatural from the bulk (or supergravity) point of view. In fact it cannot happen for canonically normalized fields.
However, it is very natural from the boundary point of view since the small number is the coefficient of the beta function which arises at the perturbative one--loop level.
We see that a potential that looks like fine tuned from the bulk point of view arises very naturally from the boundary field theory. Holography connects the parameters
of the bulk potential to parameters of the boundary theory that arise only at the one--loop level. Another fine tuning (from the bulk point of view) is the
energy scale of inflation. COBE data requires $H<10^{14}~GeV$ which may look like fine tuning since the natural energy scale for gravity is $M_P$. On the boundary field
theory this translates into the very large value of $C$ in the far IR which cannot be easily explained

We can also try to find out how ordinary inflation models look like from the boundary point of view. Consider the simplest case of chaotic inflation with
$V(\phi) =a \phi^n$[\CHA]. Then from eq. (10) we find that for the boundary theory $\beta=-n/g$ which looks like a nonperturbative result. It is easy to see that other
inflationary models with more complicated inflaton potentials will give similar results. Therefore it seems that holographic inflation with a perturbative boundary
field theory cannot reproduce any of the known inflation models.

\medskip
\centerline{\bf 4. Holographic Quintessence}
\medskip

Any model of inflation can also be a candidate for a model of quintessence with a suitable change of parameters[\QUI,\STE]. If we want to apply the above ideas to quintessence
we should first point out
the differences between quintessence and inflation. First, quintessence occurs at late times, i.e. the far UV in the boundary field theory. This means we need a model with a
UV fixed point. Second, the most severe constraints for inflation which are those coming from the number of e--foldings and the magnitude of density perturbations do not
apply to quintessence. (Of course, the slow--roll conditions $\beta, \gamma<<1$ should hold for quintessence.)
Third, since we know the value of the vacuum energy now to be $V \sim 10^{-120}M_P^4$, we know the value of the $C$--function of the field theory, $C \sim 10^{120}$
(which is inverse the vacuum energy in Planck units).

The energy density and pressure of a scalar field are given by
$$\rho={1 \over 2} (\partial_t \phi)^2+V(\phi) \eqno (24)$$
and
$$p={1 \over 2} (\partial_t \phi)^2-V(\phi) \eqno (25)$$
The equation of state is defined as $\omega=p/\rho$. Quintessence corresponds to a scalar with $-1<\omega<-1/3$ which gives rise to
an accelerating expansion of the universe. It can be shown that $\omega$ is related to the $\beta$--function of the boundary field theory as
$$3(\omega+1)=\beta^2 \eqno(26)$$
Note that for a cosmological constant $\omega=-1$ and therefore $\beta=0$. This corresponds to the description of de Sitter space by a CFT (field theory at the fixed
point). Observations suggest that $-1<\omega<-2/3$ which can be written as $0 < \beta^2 <1$. This should be satisifed by the field theory near its UV fixed point
if it is to describe quintessence. Since the slow--roll condition is $\beta<<1$ we find that the observational constraint on quintessence is automatically
satisfied by the boundary theory. Thus we find that for holographic quintessence is not constrained by observations (at least not yet).

Note that a UV fixed point
naturally leads to values of $\omega$ very close to $-1$. Consider for example
the second example in the previous section. This theory has $\beta= ag-bg^2$ and therefore a UV fixed point at $g=a/b$. If we want the field theory to
be perturbative at (or near) the UV fixed point we demand $a<b$ which for example is satisifed by the three dimensional nonlinear $\sigma$ model. Then near the fixed
point $\beta \sim (a/b)^2<<1$. Since $\omega=\beta^2/3-1$ from eq. (26) we find that holographic quintessence mimics a cosmological constant.
The quintessence potential in the bulk is again given by eq. (23) which looks very much like the models considered in [\QUI].

\medskip
\centerline{\bf 5. Conclusions and Discussion}
\medskip

In this paper we described the holographic inflation scenario which assumed a generalization of the dS/CFT correspondence to asymptotically de Sitter space--times. We found
the constraints inflation imposes (such as slow--roll, number of e--foldings and density perturbations) on the boundary field theory and showed that they can be easily realized in
principle. However, it is more difficult to find specific three dimensional Euclidean field theories that lead to inflation. Thus, construction of a working holographic
inflation model remains a challenge.

Nevertheless, the holographic inflation
scenario has some model independent predictions. The demand for a perturbative boundary theory leads to $H<10^{14}~GeV$ which is the constraint from COBE.
The correct amount of density perturbations can only be obtained if $r<0.08$ and $n_T<0.04$. Thus this scenario predicts a very small amount of tensor component of
density perturbations (compared to the scalar component) which is almost scale invariant. In addition, $n_T$ and $r$ are proportional to the value of inflationary potential;
so a lower inflation scale results in smaller $n_T$ and $r$.

We also found the form of the bulk
inflaton potential that corresponds to holographic inflation. These exponential potentials look like fine--tuned from the bulk point of view but naturally arise
in the boundary field theory. The small numbers that appear in the bulk potential originate from perturbative one--loop effects on the boundary.
One can extend this scenario to quintessence and obtain holographic quintessence. In this case, the demand for a
perturbative (UV) fixed point gives an equation of state very close to that of a cosmological constant.

The main assumption of the holographic inflation scenario is the dS/CFT correspondence. Although there is a large amount of circumstantial evidence in favor of this
duality, it has a number of serious problems as mentioned in the introduction. It would be nice if the correspondence could be made stronger so that holographic
inflation sits on firmer ground.
As mentioned above, it is easy to find three dimensional Euclidean field theories which lead to inflation in principle but not in practice. It is important to find at least
one boundary field theory that satisfies all of the constraints arising from inflation as an existence proof of the scenario. This problem may be overcome if one looks
for models of holographic inflation with
more than one field (such as in hybrid inflation[\HYB], D--term inflation[\DTE,\BIN,\TYP], P--term inflation[\REN,\MAR] and D--brane inflation[\BRA-\KAL]).
On the boundary these would correspond to theories with two coupling constants and two operators. In addition, the bulk inflaton fields are not free in these cases which
means the connection between the inflaton mass and $\gamma_O$ is not simple.
We leave these interesting but  more complicated possibilities for future work.



\vfill

\refout

\end
\bye